# Multiple Spin State Analysis of Magnetic Nano Graphene

Norio Ota*, Narjes Gorjizadeh** and Yoshiyuki Kawazoe**

*R&D Division, Hitachi Maxell Ltd., 1-1-88 Ushitora, Ibaraki-city, Osaka 567-8567 JAPAN

** Institute for Materials Research, Tohoku University., 2-1-1 Katahira, Aoba-ku, Sendai 980-8577 JAPAN

Recent experiments indicate room-temperature ferromagnetism in graphite-like materials. This paper offers multiple spin state analysis applied to asymmetric graphene molecule to find out mechanism of ferromagnetic nature. First principle density functional theory is applied to calculate spin density, energy and atom position depending on each spin state. Molecules with dihydrogenated zigzag edges like $C_{64}H_{27}$, $C_{56}H_{24}$, $C_{64}H_{25}$, $C_{56}H_{22}$ and $C_{64}H_{23}$ show that in every molecule the highest spin state is the most stable one with over 3000 K energy difference with next spin state. This result suggests a stability of room temperature ferromagnetism in these molecules. In contrast, nitrogen substituted molecules like $C_{59}N_5H_{22}$, $C_{52}N_4H_{20}$, $C_{61}N_3H_{22}$, $C_{54}N_2H_{20}$ and $C_{63}N_1H_{22}$ show opposite result that the lowest spin state is the most stable. Magnetic stability of graphene molecule can be explained by three key issues, that is, edge specified localized spin density, parallel spins exchange interaction inside of a molecule and atom position optimization depending on spin state. Those results will be applied to design a carbon-base ferro-magnet, an ultra high density 100 tera bit /inch$^2$ class information storage and spintronic devices.

## 1. Introduction

Carbon based room-temperature ferromagnetic materials are very attractive in many applications such as a light weight ecological magnet[1)-4)], an ultra high density 100 tera bit/inch$^2$ class information storage[5)] and novel spintronic devices[6)-7)]. Recently several experiments have shown room temperature ferromagnetism in modified graphite materials. Esquinazi et al.[8)] predicted ferromagnetic magnetization loop in proton ion irradiated graphite. Magnetic moments with orders of $5\times10^{-6}$ emu were observed at a room-temperature having a coercive force of 100Oe. Kamishima et al.[9)] synthesized magnetically attracted graphite-like powder. Estimated Curie temperature is very high, up to 800K, having 0.5emu/g saturation magnetization. Very recent experiment by Wang et al [10)] indicate that reduced graphene oxide shows 0.02emu/g saturation magnetization at 300K. Also, strong magnetism was observed at the defect edge of cracked graphite by Cervenka et al.[11)] Those experiments encourage us to open a new door to carbon ferromagnetism.

. There are many theoretical predictions on zigzag edge carbon magnetism due to localized density of states near Fermi energy[12)-16)] resulting antiferromagnetic feature with total magnetization of zero. Also, rectangular shaped graphene nano dot has been extensively studied[17)-20)], but only shows singlet state with zero magnetization. Whereas, Kusakabe and Maruyama[21)22)] proposed an asymmetric ribbon model showing ferrimagnetic behavior with non-zero total magnetization. This gave us a hint to analyze room-temperature ferromagnetism. Unfortunately, they did not apply to nano-meter-length graphene, and did not to analyze multiple spin states.

Reported experiments mostly use powder samples, which suggest us to apply nano-meter-size molecules. In such molecule, there appears complex multiple spin state. Questions are, which is the most stable energy states, and how about energy difference between spin states to explain the room temperature stability. In a previous paper [23)], we proposed an $C_{48}H_{28}$ molecule model with dihydrogenated zigzag edge. The first principle theory calculation shows that high spin state (Sz=3/2) is more stable than low spin state (Sz=1/2). Such results teach us the importance of multiple spin state analysis in systematic molecule group. This paper reports multiple spin state analysis in two groups. One is dihydrogenated zigzag edge group like $C_{64}H_{27}$, $C_{56}H_{24}$, $C_{64}H_{25}$, $C_{56}H_{22}$ and $C_{64}H_{23}$. The other one is nitrogen substituted molecule like $C_{59}N_5H_{22}$, $C_{52}N_4H_{20}$, $C_{61}N_3H_{22}$, $C_{54}N_2H_{20}$ and $C_{63}N_1H_{22}$. Through those analysis, we could find that high spin state of dihydrogenated molecules are very stable and suitable for realizing room-temperature ferromagnetism.

## 2, Application Example and Selection of Calculation Model

One of the important applications of nano-meter-size ferromagnetic materials is information storage. Current hard disc storage has a density around 0.7tera bit/inch$^2$ and now targeting 1 tera bit /inch$^2$ density [5] with 10 nm length, 25nm width magnetic mark. Whereas, recent scientific report predicts a possibility of single atom base storage [24] with density of $10^4$ tera bit/inch$^2$. There is a missing link between those two density regions. In order to link those two density regions, one promising candidate is a ferromagnetic molecule dot array having a typical unit size of 1 nm length, 2.5 nm width having $10^2$ tera bit /inch$^2$. Fig.1 shows a magnetic recording image using such molecule array.

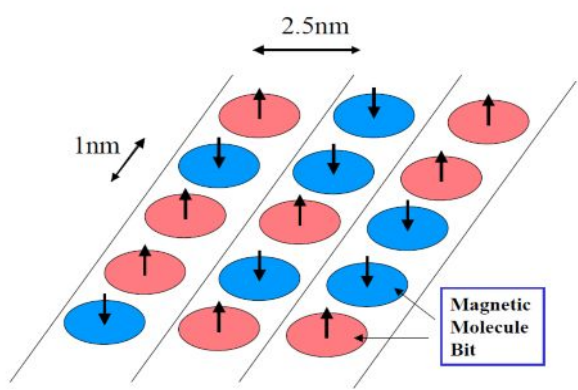

**Fig.1** Ultra high density 100 tera bit/inch$^2$ magnetic recording image using ferromagnetic molecule dot array.

For modeling such nano-meter-size molecule having asymmetric dihydrogenated zigzag edges, our calculation tried five typical examples like $C_{64}H_{27}$, $C_{64}H_{25}$, $C_{64}H_{23}$, $C_{56}H_{24}$ and $C_{56}H_{22}$ as illustrated in Fig.2. Those molecules have different numbers of dihydrogenated edges (CH$_2$-edge) from five to one. Molecule size of $C_{64}H_{27}$ is 2.16nm x 0.93nm.

In every molecule, complex multiple spin states may appear. If the highest spin state is the lowest stable energy with an energy difference of kT=1000K with the next state, we can believe room-temperature ferromagnetism.

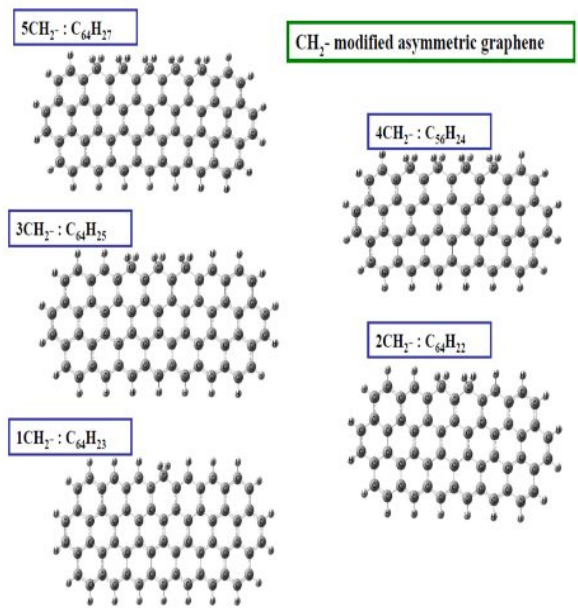

**Fig.2** Five model molecules with different numbers of dihydrogenated zigzag edges: $C_{64}H_{27}$, $C_{56}H_{24}$, $C_{64}H_{25}$, $C_{56}H_{22}$ and $C_{64}H_{23}$.

## 3. Calculation Methods

In order to clarify magnetism, we have to obtain (i) spin density map, (ii) total molecular energy and (iii) optimized atom arrangement, which are depending on every spin state. Density functional theory (DFT) [25)26] based generalized gradient approximation method (GGA-UPBEPBE) [27] is applied for those calculations. Functional basis is 6-31G [28]. Energy accuracy is required at least 10E-8 as total molecular energy after repeating atomic position optimization.

## 4, Dihydrogenated Graphene-like Molecule

In $C_{64}H_{27}$ molecule with five CH$_2$-modified edges, there are three spin state capabilities as like Sz=1/2, 3/2 and 5/2. Spin density maps at a contour surface of 0.001e/A$^3$ are shown in Fig.3, where up spin is shown in red (dark gray), down spin in blue (light gray). Looking at those figures, we can notice the following features,

(1) In every spin state, localized spin density at CH$_2$-edge shows similar feature, that is, there appears up spin on two hydrogen sites and down spin on zigzag edge carbon. Localized (CH$_2$-orbit) molecular Hund-rule plays a major role for such edge specified spin density distribution. In unlimited length ribbon case, this causes zigzag edge ferro- and ferri-magnetism[12)21)]. Advantage of molecule model is a real space representation for easier understanding of magnetism.

(2) In case of Sz=5/2, inside a molecule, up and down spins are alternately arranged one by one very regularly.
(3) Whereas, in case of Sz=3/2 and 1/2, there appears up-up and down-down spin pairs inside a molecule.

Exchange coupling between up-up (down-down) spins makes the local binding energy increase and finally elevates total molecular energy. In $C_{64}H_{27}$ molecule, total molecular energy is calculated and compared with each other. Results are shown in Fig.4. The lowest energy is obtained in the spin state Sz=5/2. Energy difference between Sz=5/2 and Sz=3/2 is 6.1 kcal/mol. This value is estimated to overcome kT=3000K as the thermal excitation energy.

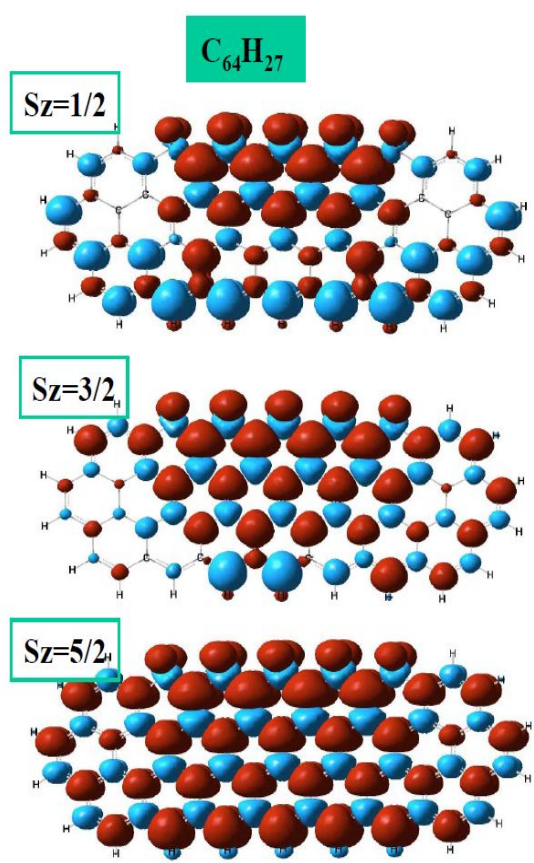

**Fig. 3.** Spin density map of $C_{64}H_{27}$ molecule having three spin states Sz=1/2, 3/2, 5/2 respectively. Red (dark gray) shows up spin, whereas blue (light gray) down spin at 0.001e/A³ spin density contour surface.

In order to obtain more systematic behavior, we changed numbers of $CH_2$- modified edges, as $C_{56}H_{24}$ (four $CH_2$- edges), $C_{64}H_{25}$ (three), $C_{56}H_{22}$ (two) and $C_{64}H_{23}$ (one). Every molecule has several spin states. Energy differences between spin states are calculated and results are shown in Fig. 4. It is very clear that in every molecule the highest spin state is the lowest and stable energy state. By these calculations, we can expect room-temperature ferromagnetism in dihydrogenated asymmetric graphene molecule.

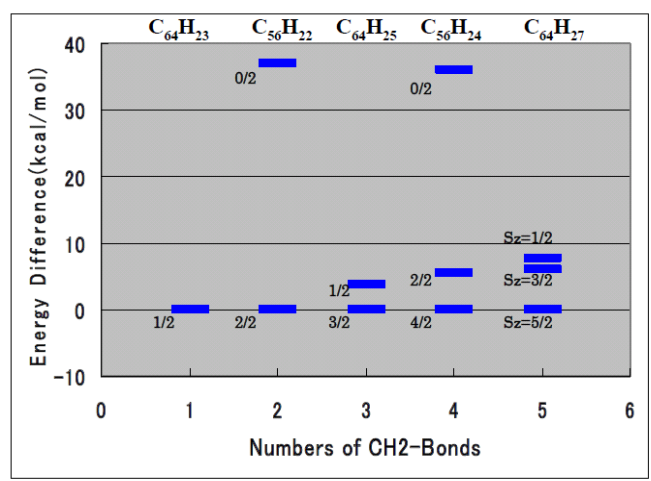

**Fig. 4.** Energy difference between spin state for $CH_2$- modified five molecules. Highest spin state is most stable in every molecule.

### 5, Nitrogen Substituted Graphene Molecule

Triggered by an experiment of K.Kamishima et al[4)], which uses nitrogen contained small molecule as a starting material, we imagined some nitrogen atoms may substitute zigzag edge carbons and bring ferromagnetism. Model molecules are shown in Fig.5, where nitrogen atoms are illustrated by blue ball.

In $C_{59}N_5H_{22}$ molecule, there are three spin states as Sz=1/2, 3/2 and 5/2. Spin density map is calculated as shown in Fig.6. Features are summarized as follows,

(1) In case of Sz=1/2, there occurs canceling of spins around nitrogen atoms (blue N marks), also there appears up and down alternative spin array around opposite side zigzag edge.
(2) Whereas in Sz=3/2, many up-up spin pairs appear on every NH- site. Also in Sz=5/2, stronger up-up spin pairs appear. These many parallel spin pairs may increase exchange interaction and finally bring unfavorable molecular energy increase

Above observation suggests the lowest and stable energy to be Sz=1/2 in NH- edge graphene molecule.

Also, we found that optimized atomic position drastically changes depending on spin state as shown in Fig.7. Right illustrations in Fig.7 indicate tilt angle θ of hydrogen atom from a molecular plane.
(1) In case of Sz=1/2, four NH- site show SP$^2$-like atomic arrangements which brings almost no spin density around top zigzag edge. Only at center NH-site, we can see SP$^3$ like atomic arrangement, which may bring up and down spin alternative arrangements on opposite zigzag edge.
  Concerning SP$^2$ positioned nitrogen, bond length between nitrogen and neighbor carbon is 0.1412nm, which is 0.7% shorter than carbon- carbon length of 0.1421nm. Angle of C-N-C is 119 degree, whereas C-C-C is 122 degree. Such slight atomic position change gives remarkable spin density change at a zigzag edge corner.
(2) In case of Sz=3/2 and 5/2, all NH- sites show SP$^3$-like arrangements at zigzag edge. Such arrangement brings strong exchange coupling between up spin on hydrogen and up spin on nitrogen.
(3) We checked CH$_2$-modified molecule case. In all molecules and every spin state we can not find any drastic change in atomic arrangements.

We added calculations on four molecules as like C$_{52}$N$_4$H$_{20}$ (four NH- edges), C$_{61}$N$_3$H$_{22}$ (three), C$_{54}$N$_2$H$_{20}$ (two) and C$_{63}$N$_1$H$_{22}$ (one).
As shown in Fig.8, energy difference between spin states is completely opposite with that of CH$_2$-cases. In every molecule, lowest energy and stable spin state is the lowest spin state to be Sz=1/2 or 0/2. We cannot explain strong magnetism by such nitrogen substituted asymmetric molecule model.

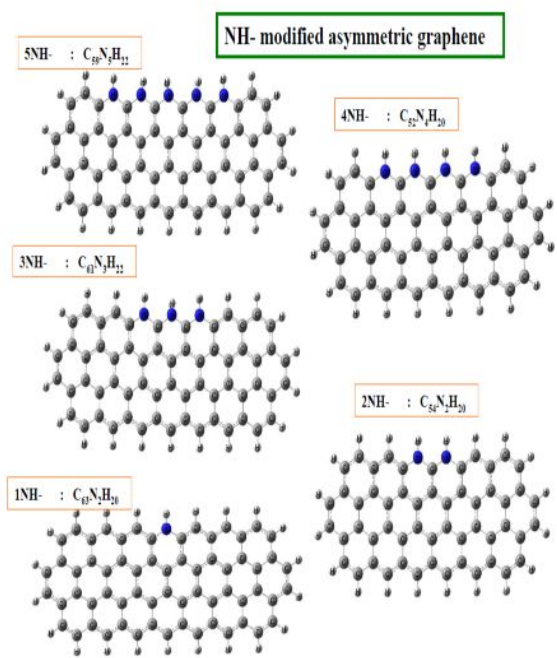

**Fig.5** Nitrogen substituted graphene molecules. Blue balls show nitrogen.

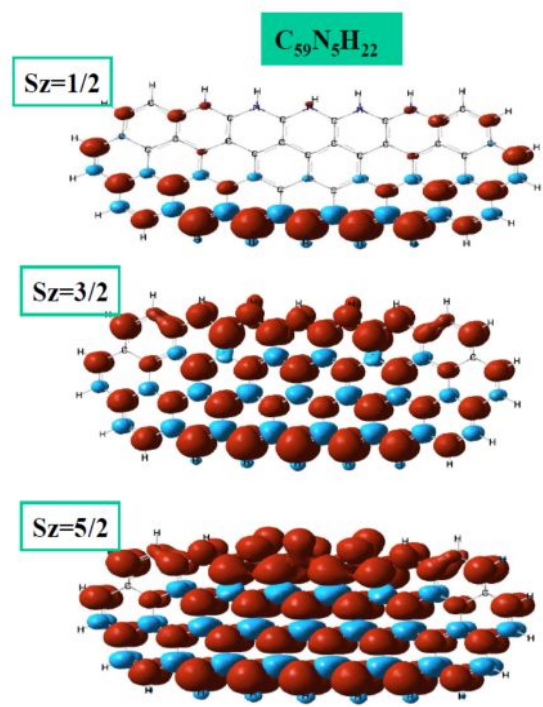

**Fig. 6.** Spin density map of C$_{59}$N$_5$H$_{22}$ molecule having three spin states Sz=1/2, 3/2 and 5/2 at a contour surface of 0.001e/A$^3$.

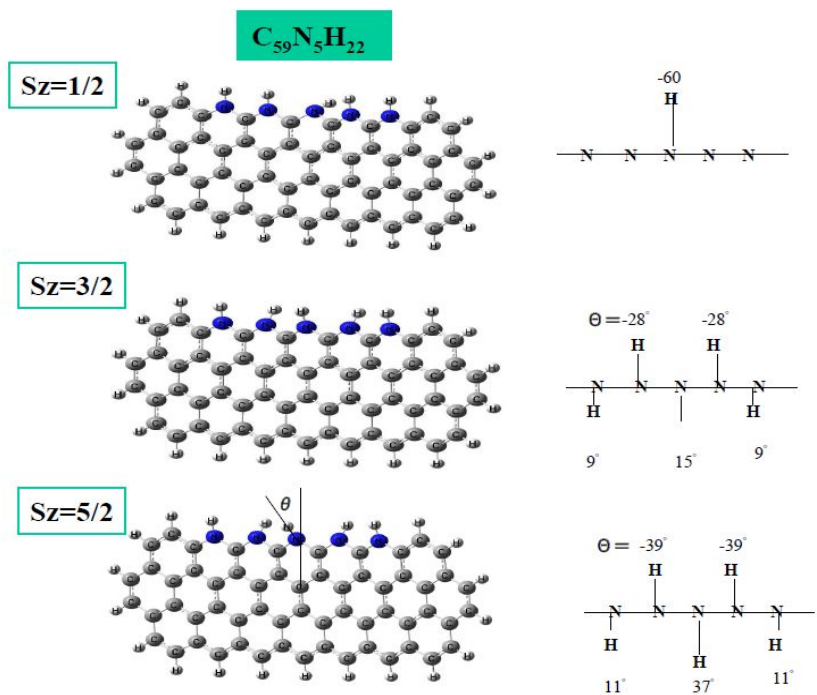

**Fig. 7.** In $C_{59}N_5H_{22}$ molecule, atoms show different arrangements depending on spin states. Left three figures show plane views observed slightly tilting from right side. Right illustrations indicate tilt angle $\theta$ of hydrogen atom bonded to nitrogen from a top view.

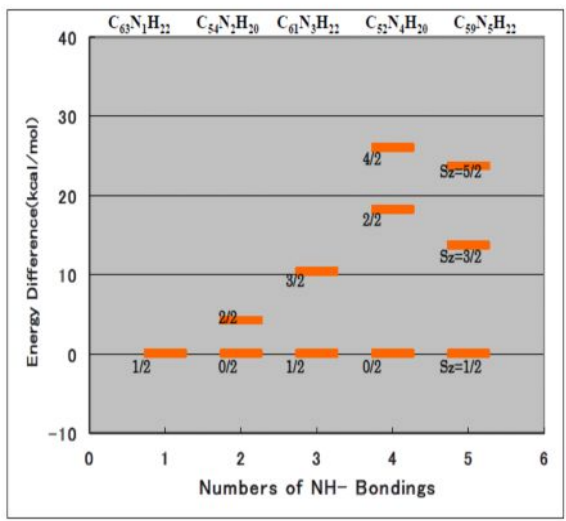

**Fig. 8** Energy difference between spin states for NH-modified five molecules. Lowest spin state show most stable in every molecule.

### 5, Conclusion

Recently, several experiments on graphite like materials show room-temperature ferromagnetism. In order to find out such mechanism, multiple spin state analysis is applied to nano-meter-size graphene-like molecules. First principle density functional theory is applied to calculate spin density distribution, molecular energy and optimized atomic position depending on each spin state.

(1) Molecules with dihydrogenated ($CH_2$-) zigzag edges like $C_{64}H_{27}$, $C_{56}H_{24}$, $C_{64}H_{25}$, $C_{56}H_{22}$ and $C_{64}H_{23}$ show that in every molecule the highest spin state is the most stable. Energy difference between the most stable spin state and the next one overcomes kT=3000K temperature difference. This result suggests stability of room-temperature ferromagnetism.

(2) In contrast, nitrogen substituted asymmetric molecules like $C_{59}N_5H_{22}$, $C_{52}N_4H_{20}$, $C_{61}N_3H_{22}$, $C_{54}N_2H_{20}$ and $C_{63}N_1H_{22}$ show the opposite result, that is, in every molecule the lowest spin state is the most stable as Sz=1/2 or 0/2. We cannot explain strong magnetism by such (NH-) zigzag edge molecule model.

(3) Magnetic stability of graphene-like molecule can be explained by following three key issues,
  1) Zigzag edge specified localized spin density,
  2) Exchange interactions between parallel spins (up-up or down-down) inside of a molecule,
  3) Atom position optimization depending on spin

state.

Those results are useful to design a carbon-base ferro-magnet, an ultra high density 100 tera bit/inch$^2$ class information storage and novel spintronic devices.

## Acknowledgements

Narjes Gorjizadeh would like to thank the crew of the Center for Computational Materials Science, Institute for Materials Research of Tohoku University for their support of the Hitachi SR11000(model K2) supercomputer system, and Global COE Program "Materials Integration (International Center of Education and Research),Tohoku University," MEXT, Japan, for financial support.